\documentstyle[twoside,epsf]{article}
\begin{document}
\hspace{7.5cm} Preprint Bu-He 93/6
\vspace{1.5cm}
\large
\bf
\begin{center}
Nuclear recoil measurements in Superheated Superconducting Granule detectors
\end{center}
\vspace{2cm}
\normalsize
M. Abplanalp, C. Berger, G. Czapek, U.~Diggelmann, M. Furlan,
A.~Gabutti, S. Janos, U. Moser, R. Pozzi, K. Pretzl, K. Schmiemann \\
{\it Laboratory for High Energy Physics,  University of Bern, Sidlerstrasse 5,
CH 3012 Bern, Switzerland\\}
D. Perret-Gallix\\
{\it LAPP, Chemin de Bellevue, 74941 Annecy, France\\}
B. van den Brandt,J.A. Konter, S. Mango\\
{\it Paul Scherrer Institute, CH-5232 Villigen PSI, Switzerland}
\vspace{2cm}
\rm
\begin{center}
\section*{Abstract}
\end{center}
The response of Superheated Superconducting Granule (SSG)
devices to nuclear recoils has been explored by  irradiating SSG
detectors with a 70Me$\!$V neutron beam. In the past we have tested Al SSG and
more recently, measurements have been performed with Sn and Zn detectors.
The aim of the experiments was to test the sensitivity of SSG detectors to
recoil energies down to a few ke$\!$V.
In this paper, the preliminary results of
the neutron irradiation of a SSG detector made of Sn
granules 15-20$\mu$m in diameter will be discussed. For the first time,
recoil energy thresholds of $\sim$1ke$\!$V have been measured.
\footnote{Talk held at the Fifth International Workshop on Low Temperature
Detectors in San Francisco, July 28th - August 3rd 1993}

\section{INTRODUCTION}
Superheated Superconducting Granule detectors (SSG) are being presently
developed for dark matter detection \cite{dmberk}.
Weakly interacting massive particles (WIMPs)
can be detected measuring the recoil energy released when they interact
with a nucleus inside the granule via  neutral-current scatterings.
A review of the
status of the SSG detector development can be found in
\mbox{Ref.\cite{Pretzl}}.
The sensitivity of SSG detectors to minimum ionizing particles
\cite{mips} and to x-rays \cite{xray} has been proven in the past. To
study the response of the detector to nuclear recoils in the ke$\!$V range,
a set of experiments have been performed by our group irradiating SSG
detectors  with a 70Me$\!$V neutron beam at the Paul Scherrer Institute in
Villigen (Switzerland).
Due to the fast transition time of the
granules \cite{Miha}, coincidences between the scattered neutrons and the
SSG  were clearly established making irradiation tests of SSG
a powerful probe of the response to nuclear recoils of superconducting
materials.

The results of the irradiation test of a SSG detector made of Al
granules 20-25$\mu$m in diameter are discussed in \mbox{Ref.\cite{neutron}}.
More recently (June 1993) we have performed measurements with
Zn (28-30$\mu$m) and Sn (15-20$\mu$m) SSG detectors.
In this paper, the preliminary results of the neutron irradiation test of
the Sn SSG detector  will be discussed.
The measured recoil energy distributions will be compared
with Monte Carlo simulations, showing that the sensitivity of Sn SSG
detectors to nuclear recoils is approximately two times higher than the
theoretical expectations of the global heating model.
Energy thresholds of $\sim$1ke$\!$V were reached in the Sn SSG.
Due to the limited angular resolution of the neutron
detector we were not able to test the SSG sensitivity to lower recoil
energies.

\section{EXPERIMENTAL SETUP}
\label{chapter2}

The SSG detector consisted of a hollow Teflon
cylinder (4mm inner diameter and 8mm inner length) filled with Sn granules
imbedded in an Al$_{2}$O$_{3}$ granulate with a volume filling factor of 15\%.
The SSG target was surrounded by a pickup coil with 180 windings connected to
a J-FET preamplifier working at room temperature.
To increase the statistics we used two identical detectors made of Sn
granules 15-20$\mu$m in diameter. Each target
consisted of $\sim$5$\cdot$10$^{6}$ granules.
The SSG detectors were operated in the mixing chamber of a dilution
refrigerator \cite{cryo} at a temperature of 40mK.

The neutron beam was generated by irradiating a beryllium target with
72Me$\!$V protons, and the residual protons were swept away by a bending magnet
downstream of the Be target. The produced neutrons were sharply peaked at
70Me$\!$V and had a  repetition rate of 17MHz
and a bunch width of 2ns \cite{Henneck}.
The beam was collimated down to a diameter of 3.2mm, in line with the SSG
detectors. The scattered neutrons were detected using a scintillator hodoscope
consisting of 18 elements placed 2 meters downstream of the SSG
detectors. The range of scattering angles covered by the hodoscope was
0.02-0.5rad with a resolution of 20mrad FWHM.

We improved the setup used in the previous irradiation tests
\cite{neutron} using concrete blocks covering the sides and the top
of the hodoscope to shield against neutrons and charged particles
background. In addition, the neutron beam flux was measured using a thin
$CH_{2}$ target positioned after the collimator and a telescope at 13
degrees to detect protons coming from $n-p$ reactions in this target.
This allows an absolute evaluation of the SSG detection efficiency.
To discriminate against charged particles entering or leaving the SSG,
two $5mm$ thick scintillator veto counters were mounted before and after the
cryostat window. The energy threshold of the counters was 0.15Me$\!$V. A third
2cm thick scintillator counter was located in front of the hodoscope.

\section{MEASUREMENTS}
\label{chapter3}
The detector sensitivity to elastically scattered neutrons was measured at
different detector thresholds.
To select a neutron induced event within the detector, coincidences in time
between the injector radio frequency, the SSG and the hodoscope signals were
established.
The measured distribution of the time difference between the Sn SSG and the
hodoscope signals is shown in \mbox{Fig. 1$a$} . Coincidences
in time between the SSG and the hodoscope signals were clearly established.
The standard deviation  of the distribution of the coincident SSG signals above
background is 25ns.
In the data analysis, only the coincidences within the hatched
region were considered.  The events outside this region were used to evaluate
the accidental background which was subtracted from the selected sample after
normalization.
A further cut on the scattered neutrons time of flight of $\pm$1ns was
introduced in order to select elastic scattering events.  The
resolution was dominated by the bunch width of the beam, and therefore the
accuracy in the energy measurement of the neutrons was $\pm$8Me$\!$V.
An elastically scattered neutron produces the phase transition (flip) of a
single granule while charged particles can cause more than 1
granule to flip.
The multiplicity spectrum of the flips in the Sn SSG is shown in
\mbox{Fig. 1$b$}. These measurements were taken irradiating
the SSG without coincidence requirements.
 Single flip events  can be clearly distinguished from events with higher
multiplicity.
To reduce the background from charged particles entering the
SSG and from charge exchange reactions within the SSG detector,
only events with no signal from the veto counters and with a single flip
multiplicity were considered.

\section{RECOIL ENERGY THRESHOLD IN SSG}
\label{chapter4}

The energy threshold $E_{th}$ of a single
granule is given by the energy needed to rise the granule temperature from
the bath $T_{b}$ to the transition $T_{sh}$ temperature. In the global
heating model, the energy threshold of a granule of volume V is:
	\begin{equation} \label{eq:Eth}
		E_{th}=V \cdot \int_{T_{b} }^{T_{sh}} C(T) \; dT
	\end{equation}
where {\it C(T)} is the superconducting specific heat \cite {specificheat}.
{}From the phase diagram, the change in temperature needed to produce the
phase transition
can be related to the magnetic threshold $h=1-H_{a}/H_{sh}$ with
$H_{sh}$ the granule superheating field and $H_{a}$ the applied field
strength.
The calculated energy thresholds of 17$\mu$m Sn, 22$\mu$m Al and 30$\mu$m Zn
granules are
compared in \mbox{Fig. 2} at the temperature of 40mK.

The  recoil energies due to
neutral-current scatterings of dark matter particles in Sn detectors
are expected to be of the order of a few ke$\!$V  \cite{Gabutti}.
Such thresholds can be reached with 17$\mu$m Sn granules at magnetic
thresholds of h$\sim$0.01.

To evaluate the SSG sensitivity to nuclear recoils the irradiation
measurements were compared to Monte Carlo simulations using the same procedure
as discussed in \mbox{Ref.\cite{neutron}}.
The scattering angle distribution of elastically scattered neutrons within
the SSG was obtained by a partial wave expansion using the optical model
\cite{opticalmodel}. Each SSG phase transition was simulated considering
one granule randomly selected from a pool of granules with the same
distributions of individual flipping fields and sizes of the SSG detector.
The recoil energy $E_{r}$ deposited inside the granule was determined from
the scattering angle $\theta$ using the kinematical condition:
	\begin{equation} \label{eq:Erec}
E_{r}=4 \cdot sin^{2}(\frac{\theta }{2}) \cdot \frac{M_{n}}{M} \cdot E_{n}
	\end{equation}
where $E_{n}$ is the neutron energy and $M_{n}$ and $M$ are the masses of
the neutron and of the target nucleus respectively. The value of $E_{r}$ was
then compared to the expected energy threshold of the granule.
In \mbox{Fig. 3}
the scattering angle distributions measured with the Sn SSG at the
thresholds $h$=0.01 and $h$=0.025 are compared to the calculations.
The measured distributions are normalized to the neutron flux. The
simulated distributions are normalized to the measured distributions.
The first (0.15rad) and the second (0.4rad)  diffraction maximum
are well visible in the measured scattering angle distributions.
The measurements show the expected shift towards higher scattering
angles when the detector threshold is increased.

Previous measurements on SSG detectors,
irradiated with minimum ionizing particles \cite{mips}, have shown
that Sn granules are a factor of two more sensitive than expected from the
global heating model (equation \ref{eq:Eth}).
To investigate such behaviour, the measured recoil energy distributions
were compared to Monte Carlo simulations which included in one case global
heating and in the other case a factor of two higher sensitivity.
The comparisons are shown in \mbox{Fig. 4} using the same
normalization as in \mbox{Fig. 4}. In the final data analysis
the absolute detector efficiency will be evaluated.
The data in \mbox{Fig. 4} seem to indicate that Sn granules
are more sensitive to nuclear recoils than one would expect from the global
heating model.  The increased sensitivity may be due to local heating
effects.

The cutoff in the measured  recoil energy distribution at $h$=0.035 is sharp
and corresponds to an energy threshold of 5ke$\!$V. At the lower threshold
$h$=0.025, the onset of the measured recoil energy distribution
is at $\sim$2ke$\!$V.  At the threshold $h$=0.01, the shape of the
distribution at small recoil energies is dominated by the
limited angular resolution of the neutron hodoscope (0.02rad)
preventing us to measure the sensitivity of the SSG to recoil energies
below 1ke$\!$V.

\section{CONCLUSIONS}

The sensitivity of SSG to nuclear recoils has been explored irradiating SSG
detectors with a 70Me$\!$V neutron beam.  We have proven, for the first time,
that SSG detectors made of 15-20$\mu$m Sn granules are sensitive down to
1ke$\!$V recoil energies.
The neutron irradiation experiments show, in agreement with
previous measurements with minimum ionizing particles,
that the sensitivity of Sn granules is a factor of two higher than expected
from the global heating model.
The achieved sensitivity to low recoil energies is encouraging for the use of
SSG as a dark matter detector.

\section*{Acknowledgments}

We would like to thank K. Borer, M. Hess, S. Lehmann, L. Martinez,
F. Nydegger, J.C. Roulin and H.U. Schutz from the High Energy
Physics Laboratory of the University of Bern for technical support.
This work was supported by the Schweizerischer Nationalfonds zur
Foerderung der wissenschaftlichen Forschung and by the Bernische Stiftung
zur Foerderung  der wissenschaftlichen Forschung an der Universitaet Bern.


\section*{Figure captions}
\begin{enumerate}
\item $a)$ Distribution of the time difference between the SSG and the
hodoscope signals for the Sn detector at $h$=0.01 $b)$ Multiplicity spectrum of
the Sn SSG when irradiated with the neutron beam at $h$=0.005; these
events were recorded without coincidence requirements.
\item Calculated energy thresholds [ke$\!$V], in the global heating model,
of 17$\mu$m Sn, 22$\mu$m Al and 30$\mu$m Zn  granules at 40mK.
\item Calculated (histogram) and measured (points) scattering angle
distributions in the SSG detector made of 15-20$\mu$m Sn granules at the
thresholds $a)$ $h$=0.01 and $b)$ $h$=0.025.
\item Measured (points) and calculated (histogram) recoil energy
distributions in the Sn SSG detector at the thresholds $a)$ $h$=0.01,
$b)$ $h$=0.025 and $c)$ $h$=0.035. The calculated distributions at the $(
bottom$) are derived
from the global heating model and the ones at the $(top)$  using a factor of
two higher sensitivity.
\end{enumerate}
\end{document}